\RequirePackage{ifpdf}
\documentclass[12pt,letterpaper]{article}
\pdfoutput=1
\usepackage{jheppub}
\usepackage{epsfig}
\usepackage{enumitem}
\usepackage[latin1]{inputenc}
\usepackage{bbm,amsfonts}
\usepackage{rotating,graphicx}
\usepackage{amssymb,amsmath,amsfonts,mathtools}
\usepackage{fancybox,diagbox}
\usepackage{enumerate}
\usepackage{dsfont}
\usepackage{verbatim}
\usepackage{wrapfig}
\usepackage{slashed}
\usepackage{shuffle}
\usepackage{subcaption}
\usepackage{braket}
\usepackage{pdflscape}
\usepackage{accents}
\usepackage{afterpage}

\author[a]{Marco S. Bianchi}
 
\affiliation{Instituto de Ciencias F\'isicas y Matem\'aticas, Universidad Austral de Chile, Casilla 567, Valdivia, Chile}

\emailAdd{marco.bianchi@uach.cl}  
  
\abstract{I consider three-point functions of twist-two operators in ${\cal N}=4$ SYM, two of which endowed with spin. I supply perturbative data up to twelve units of spins and second perturbative order at weak coupling.
}

\title{Three twist-two, two spins, two loops.} 

\newcommand{\be}{\begin{equation}}
\newcommand{\ee}{\end{equation}}
\newcommand{\beq}{\begin{equation}}
\newcommand{\eeq}{\end{equation}}
\newcommand{\bea}{\begin{eqnarray}}
\newcommand{\eea}{\end{eqnarray}}
\newcommand{\ena}{\end{eqnarray}}

\def\Tr{\textrm{Tr}}

\numberwithin{equation}{section}

\def\clock{{\count0=\time
           \divide\count0 60
           \ifnum\count0<10 0\fi\the\count0
           \multiply\count0 -60 \advance\count0 \time
           :\ifnum\count0<10 0\fi \the\count0
         }}
\newcommand{\timestamp}{{\small\vbox{\hbox{\tt\jobname.tex}
\hbox{\the\day/\the\month/\the\year, \clock}}}}
\newlength{\dhatheight}
\newcommand{\doublehat}[1]{%
    \settoheight{\dhatheight}{\ensuremath{\hat{#1}}}%
    \addtolength{\dhatheight}{-0.35ex}%
    \hat{\vphantom{\rule{1pt}{\dhatheight}}%
    \smash{\hat{#1}}}}

\begin{document}

\maketitle
\allowdisplaybreaks

\section{Introduction}

As a conformal field theory, $\mathcal{N}=4$ SYM is defined by the spectrum of its operators and their structure constants.
Both problems can be attacked, in principle, with integrability \cite{Beisert:2010jr,Basso:2015zoa}.
This note focuses on structure constants. In particular, it addresses the computation of three-point functions involving more than one operator with spin.
From the perspective of the operator-product expansion (OPE) \cite{Dolan:2001tt,Dolan:2004iy}, such coefficients emerge from a four-point function involving one un-protected operator, or alternatively from a multiple OPE of a higher-point function of protected ones, such as a five-point correlator.
Beyond one-loop order, the computation of these correlators is not as well developed as for four operators, and mostly limited to the work \cite{Eden:2010zz}.
Recently, the large spin of such higher-point correlators has been analyzed \cite{Bercini:2020msp,Bercini:2021jti}. This establishes the exact behaviour of structure constants with various spinning operators at all loops, in the large spins limit, and a duality with null Wilson loops.

The computation presented here lies in the opposite regime, that is small spins. The calculation is performed perturbatively at weak coupling. In this setting, high values of the operators spins are a nuisance rather than a blessing (ten units of spin already means high in this article). In fact, they are the main source of complexity and computational bottlenecks, which this work aims to attack.
Being the simplest, twist-two operators are considered. Their spectrum is well-known from both explicit computation \cite{Kotikov:2003fb,Kotikov:2004er} and integrability \cite{Staudacher:2004tk,Eden:2006rx,Belitsky:2006av,Beisert:2006ez}. 
Their three-point functions with two protected operators are also known to vertiginous perturbative precision \cite{Eden:2015ija,Basso:2015eqa,Eden:2016aqo,Goncalves:2016vir,Basso:2017muf,Georgoudis:2017meq}. Such results have been derived mostly from the OPE expansion of four-point correlation functions of protected operators.

Less is known for structure constants involving more than one operator with spin.
This note intends to address such a paucity of data.
The problem has already been considered in \cite{Bianchi:2019jpy}. The present work is a non-trivial extension of that.
The idea of \cite{Bianchi:2019jpy,Bianchi:2018zal} consists in tackling the computation employing an integration trick, first proposed in \cite{Plefka:2012rd}. This can be interpreted as a soft limit in momentum space. The bottom-line (more details follow in the next section) is that it allows to trade the full three-point function for a much simpler two-point-function-like object. Such a simpler computation is sufficient for isolating the structure constant of the three-point function. The most striking aspect of the simplification lies in the computation of integrals. This is performed through reduction via integration-by-parts (IBP) identities. The latter is a well-developed topic in literature, especially so for this class of propagator integrals.
Conformal invariance is instrumental in the extraction of the structure constants. It completely fixes the space-time structure of the correlator. The only dynamical information lies in its coefficient, the structure constant. This is one number at each perturbative order.
Subtleties arise, due to regularization, which will be addressed momentarily.

The punchline of \cite{Bianchi:2019jpy} is as follows.
At one loop the computation is feasible up to high values of the spins, but still in a case-by-case analysis. A solution of the IBP reduction of the relevant integrals for generic spins would grant a complete answer, but was not attempted in \cite{Bianchi:2019jpy}.
Instead, the amount of data allowed for conjecturing a heuristic formula establishing the dependence of the structure constants on spins and polarization, though in terms of some undetermined coefficients. 
I was not able to progress further and fix such remaining coefficients. More insightful people managed to do it \cite{Bercini:2021jti}.
Therefore, the problem of computing such structure constants at one loop has been completely solved by \cite{Bercini:2021jti}. 

At the next perturbative order a few results were computed in \cite{Bianchi:2019jpy}, demonstrating that they can be attacked with the method proposed.
However, the computation stopped short at rather disappointingly low values of the spins, namely six. This was due to computational complexity, the main bottleneck being integral reduction. Yet, the limitation stemmed mostly from the naivety of the approach taken for completing such a reduction: easy to implement, but too inefficient for attacking a complicated problem. 
The aim of this paper is to advance further in the two-loop computation and work out a reasonable amount of perturbative data with modest computational resources and time.
Thanks to various technical optimizations with respect to \cite{Bianchi:2019jpy}, a few-weeks computation on an ordinary laptop (plus some aid from a cluster) produced results up to twelve units of operators spins, which are presented here.
Such data are too scarce to conjecture an analogous formula to that encountered at one loop. I was hoping to find one at low values of polarizations, where the one-loop formula simplifies considerably, but had no success.
Still, this note provides some new solid results, which can be compared with alternative computation methods, such as OPE of higher-point functions and integrability.

\section{The perturbative computation}\label{sec:method}

The setting of the calculation is the same as in \cite{Bianchi:2019jpy} and I refer to that for further details.
The basic definitions of operators are reported in the appendix.
This note focuses on three-point functions with two spinning operators. In a conformal field theory their form is fixed \cite{Sotkov:1976xe}
\begin{equation}\label{eq:3ptstructure}
\left\langle \hat {\cal O}_{j_1}(x_1)\, \, {\cal O}_{BPS}(x_2) \,\,\doublehat{{\cal O}}_{j_2}(x_3) \right\rangle = \frac{ \displaystyle
\sum_{l=0}^{\text{min}(j_1,j_2)}\, {\cal C}_l\, \frac{\hat Y_{32,1}^{j_1-l}\, \doublehat{ Y}_{12,3}^{j_2-l}}{\left(x_{13}^2\right)^l} I_{13}^l}{|x_{12}|^{\Delta_{12,3}-j_1+j_2} |x_{23}|^{\Delta_{23,1}+j_1-j_2} |x_{13}|^{\Delta_{31,2}-j_1-j_2}}
\end{equation}
where $j_1$ and $j_2$ are the operators spins. The quantities $Y$ and $I$ appearing on the right-hand side are certain invariants whose precise form is spelled out in the appendix. The powers of squared distances are combinations of the conformal dimensions of the operators. The hats symbols denote contractions of the tensor structures with two distinct sets of light-cone vectors.
The index $l$ labels the polarization of the three-point functions and to each value $0\leq l\leq \text{min}(j_1,j_2)$ an independent structure constant is attached.
The objective is the two-loop computation of such structure constants.

They are extracted integrating both sides of \eqref{eq:3ptstructure} on one of the integration points of the operators.
The left-hand-side is expanded perturbatively in Feynman diagrams.
At that level, the integration translates into a soft limit in momentum space. This effectively collapses the three-point function onto a two-point function.
The latter, being much simpler than the original three-point problem, can be calculated efficiently, leveraging vast literature and techniques for dealing with two-point integrals.
In practice, the soft limit introduces additional powers of some propagators.
Conversely, the spins of the operators translate into powers of momenta.
Both occurrences are dealt with through integration-by-parts identities (IBP) \cite{Chetyrkin:1981qh,Tkachov:1981wb}, reducing all integrals to a finite set of known master integrals. 
This is the crucial technical step, allowing for carrying out the computation. More details on the procedure are articulated in the next section.

On the right-hand side of \eqref{eq:3ptstructure}, the integration over the insertion point of an operator was evaluated in general, as reported in the appendix \eqref{eq:integrated}.
After doing that, \eqref{eq:3ptstructure} allows for the extraction of the desired coefficient $\mathcal{C}$, by comparison of both sides.
This is the punchline of the calculation method.

There are a few subtleties attached to this seemingly straightforward process.
All arise due to the necessity of regularizing divergent intermediate quantities.
There are various sources of singularities. 
First, UV divergence appear. These stem physically from the two-point functions of operators and ought to be renormalized away. In fact, this provides a powerful check for the self-consistency of the computation.
Secondly, the soft limit enforced to simplify the computation introduces IR divergences. These are independent of the former and combine with them. Third, individual Feynman diagrams could possess spurious divergences which cancel away from the final result.
A regulator is needed for carrying out the computation in practice.
Dimensional regularization is used.
The advantage of dimensional regularization consists mostly in being compatible with the IBP reduction of integrals, which are carried out in arbitrary dimensions.
Moreover this is the regularization method which is usually deployed for renormalization of UV divergences.
A different regulator might be employed for IR divergences. A mass regulator for instance could regulate such singularities and separate them clearly from the former.
This would introduce a mass in the relevant integrals, hampering the computation considerably. Overall, I reckon dimensional regularization is the only viable option.

In dimensional regularization, IR and UV poles mix and multiply. After renormalizing operators, only IR divergences can  be present, arising from the soft limit.
Such divergences can behave rather heterogeneously depending on the conformal weights of the operators. 
Two qualitatively different scenarios may occur. The result may display a divergent behaviour with pole powers increasing with perturbative order; or it can present a fixed lower bound for $\epsilon$ powers at each perturbative order.
The same pattern must appear integrating the generic, conformal form of the three-point function on the right-hand-side of \eqref{eq:3ptstructure}.
Since that integral might in general be divergent, it should be regularized with the same method as in the perturbative expansion of the left-hand-side, i.e.~dimensional regularization, for consistency. 
If the order of $\epsilon$ poles is fixed for all perturbative orders, the coefficient of such poles emerges from the $d=4$ limit of the integrated expression of the three-point function. This in turn means that only the known expression of the-point function in $d=4$ is required. Any subleading in $\epsilon$ corrections can be safely discarded. The extraction of the structure constant is then correct.
On the contrary, if the order of the divergences is not fixed, there is no sensible limit to $d=4$ which can be taken. The only possible comparison between the two sides of the equation would involve considering the three-point function in $d=4-2\epsilon$ dimensions, whose form is not known in general.
This impedes a sensible extraction of the structure constant.

In the problem of this paper, integrating on the insertion point of a protected operator or of a spinning one illustrates this distinction.
The former yields a result with a constant $1/\epsilon$ pole, from which the structure constant can be extracted.
The latter produces an output with increasing powers of $\epsilon$ poles, except for the null polarization case, where the order is constant.
Indeed for null polarizations a structure constant can be extracted and is equal to the other integration result.
For other polarizations it is just impossible to disentangle the perturbative corrections of the structure constant and spurious terms due to undetermined subleading in $\epsilon$ corrections to the three-point function hitting a higher order IR $\epsilon$ pole. In fact, a naive comparison fails.
In conclusion there is only one sensible choice of the integration point for these structure constants, namely that of the protected operator.

More in general, regulating both sides of \eqref{eq:3ptstructure} introduces an order-of-limits issue.
On the right-hand-side, using the conformal expression of the three-point function entails taking the $d\to4$ limit first. Then the soft limit is taken and dimensional regularization is again used to regulate the integration over an operator insertion point.
On the left-hand-side, the soft limit is taken before on the Feynman diagrams, where dimensional regularization is applied. Only at the end of the perturbative computation the limit $d\to4$ is enforced.
I could not find any argument why the two limits should commute.
In fact, it was shown in \cite{Bianchi:2020cfn} that in general they do not, by an explicit example.
Luckily, it so happens that the present computation seems not to be affected by this issue.
This is easy to verify at one-loop order from direct inspection of the integrals.
A posteriori, this is confirmed by the fact that the one-loop structure constants computed in \cite{Bianchi:2019jpy} coincide with those determined in \cite{Bercini:2020msp,Bercini:2021jti} using an independent method.
At two loops the same analysis is difficult due to the complexity of the calculation.
Again a posteriori, the method reproduces correctly the known results for three-point functions with one spinning operator \cite{Eden:2012rr}.
Since there are no qualitatively new diagrams and integrals associated to the computation with two spinning operators (only more momentum powers), I assume the order-of-limits issue to be absent and proceed.


\subsection{Integrals treatment}

The problem described above boils down to the reduction of various integral topologies with powers of momenta in their numerators.
The latter are contracted with two different sets of null vectors $z_1$ and $z_2$. This gives rise to polynomials in the scalar product $z_1\cdot z_2$, with degree up to the minimum between the powers of $z_1$ and $z_2$ contractions with loop momenta.
At two-loop order three-loop momentum integrals are needed. 

\paragraph{IBP reduction.}
The reduction can be performed in various ways.
The integrals can be IBP reduced including directly $z_1$ and $z_2$ contractions. This introduces new external null momenta $z_1$ and $z_2$, with their respective IBP identities.
The advantage of such an approach is that it is directly implementable on IBP reducers on the market. In particular, I used FIRE6 \cite{Smirnov:2008iw,Smirnov:2019qkx} with LiteRed solved rules \cite{Lee:2012cn,Lee:2013mka} (which are more effective than Laporta reduction).
The reduction works fine and fast for two-loop integrals and for three-loop integrals, up to certain powers. 
The disadvantage is that FIRE6 tends to crash (at least on my computer) for high enough powers of numerator momenta of some three-loop integrals. Not surprisingly, the onset of this behaviour seems to happen for lower powers of momenta, when including $z_1$ and $z_2$ dependence and their additional IBP rules, than the case of integrals with no such external momenta contractions.
In particular the reduction fails for some integrals involved in the three-point function with two spin-6 operators. That is the main reason why the earlier computation of \cite{Bianchi:2019jpy} stopped short around this complexity level.

\paragraph{Tensor reduction} An alternative approach consists in performing a reduction of tensor integrals to scalar ones, later projecting the numerator onto the $z_1$ and $z_2$ contractions.
This process generates scalar integrals involving only products with the external momentum $p$. These are more easily and rapidly reduced via IBP reductions, than the aforementioned case with additional external momenta.
In this case another bottleneck is the tensor reduction itself. The computational load can be alleviated considering the symmetry properties of both numerator momenta and of the resulting contraction with null vectors. This is however a case-by-case analysis.
In order to speed up such a reduction I used the following method.
Each numerator structure of momenta is defined by the set of indices to be contracted with $z_1$ or $z_2$ factors. Let's say that there are $n_1$ contractions with $z_1$ and $n_2$ contractions with $z_2$.
The indices to be contracted with $z_1$ and $z_2$ are simmetrized (and traceless) by construction. Therefore the reduction corresponds to a case with symmetric numerator powers
\begin{equation}
k_{(\alpha_1)}\cdot z_1\,\, k_{(\alpha_{n_1})}\cdot z_1\,\, k_{(\beta_1)}\cdot z_2\,\, k_{(\beta_{n_2})}\cdot z_2\quad \rightarrow\quad k_{(\alpha_1)}^{\big\{\mu_1}\dots k_{(\alpha_{n_1})}^{\mu_{n_1}\big\}}\,\,
k_{(\beta_1)}^{\big\{\nu_1}\dots k_{(\beta_{n_2})}^{\nu_{n_2}\big\}}
\end{equation}
where the indices $\alpha$ and $\beta$ identify loop momenta in the integral.
A generic ansatz for the right-hand-side of such a tensor integral is of the form
\begin{align}
&\int\, k_{(\alpha_1)}^{\big\{\mu_1}\dots k_{(\alpha_{n_1})}^{\mu_{n_1}\big\}}\,
k_{(\beta_1)}^{\big\{\nu_1}\dots k_{(\beta_{n_2})}^{\nu_{n_2}\big\}}\, \prod\, \text{propagators} =\\
&= \sum_{i=0}^{\lfloor \frac{n_1}{2}\rfloor}\sum_{j=0}^{\lfloor \frac{n_2}{2}\rfloor}\sum_{k=0}^{\lfloor\frac{n_1-2i+n_2-2j}{2}\rfloor}\!\!\!\! c(n_1,n_2,\{\alpha\},\{\beta\})_{i,j,k}\, \bigg(g^{\mu_{1}\mu_{2}}\dots g^{\mu_{i-1}\mu_{i}} \, g^{\nu_1\nu_2}\dots g^{\nu_{j-1}\nu_j}g^{\mu_{i+1}\nu_{j+1}}\dots\nonumber\\&~~~~~~~~
\dots g^{\mu_{i+k}\nu_{j+k}}\,\, p^{\mu^{i+k+1}}\dots p^{\mu_{n_1}}\, p^{\nu_{j+k+1}}\dots p^{\nu_{n_2}}  + \text{all perms among $\mu$ and $\nu$ indices}\bigg)\nonumber
\end{align}
with the same coefficient for various tensor structures, as indicated, thanks to symmetry. Eventually the permutations all evaluate to the same result after contractions with the null vectors, giving rise to a combinatorial factor. 
The important information resides in the coefficients $c$, which depend on the particular integral.
They are indexed according to the number of different possible contractions among $z_1$- and $z_2$-to-be-contracted indices and mixed ones. These are labeled by $i$, $j$ and $k$ in the above formula.
Only the coefficients $c(n_1,n_2)_{0,0,k}$ survive the contraction with null vectors.

The system can be inverted, in principle, after multiplying both sides with suitable tensor structures.
The inversion coefficients are by construction polynomials in $z_{12}$ and rational functions of $d$, which multiply powers of external momenta and metric tensors, of a similar structure, eventually to be contracted with the original numerator momenta of the integral.
Hence, each coefficient $c$ reads
\begin{align}\label{eq:pv}
&c(n_1,n_2,\{\alpha\},\{\beta\})_{i,j,k} = \sum_{i=0}^{\lfloor \frac{n_1}{2}\rfloor}\sum_{j=0}^{\lfloor \frac{n_2}{2}\rfloor}\sum_{k=0}^{\lfloor\frac{n_1-2i+n_2-2j}{2}\rfloor}\, 
d(n_1,n_2)_{i,j,k}^{l,m,n}
\nonumber\\& 
\, \bigg(g^{\rho_{1}\rho_{2}}\dots g^{\rho_{i-1}\rho_{i}} \, g^{\sigma_1\sigma_2}\dots g^{\sigma_{j-1}\sigma_j}g^{\rho_{i+1}\sigma_{j+1}}\dots g^{\rho_{i+k}\sigma_{j+k}}\nonumber\\&~~~~~~~~
 p^{\rho^{i+k+1}}\dots p^{\rho_{n_1}}\, p^{\sigma_{j+k+1}}\dots p^{\sigma_{n_2}}  + \text{all perms among $\rho$ and $\sigma$ indices}\bigg)
\nonumber\\&
\int \, k_{(\alpha_1)}^{\rho_1}\dots k_{(\alpha_{n_1})}^{\rho_{n_1}}\,
k_{(\beta_1)}^{\sigma_1}\dots k_{(\beta_{n_2})}^{\sigma_{n_2}}\, \times \text{propagators}
\end{align}
where the permutations take care of the symmetry of numerator indices.
The inversion coefficients $d(n_1,n_2)_{i,j,k}^{l,m,n}$ multiply tensor structures contracting indices in the numerator of the to-be-reduced integral. Such structures are generic for given $n_1$ and $n_2$ and do not depend on the specific numerator or integral topology. Therefore they can be computed once and recycled into other integrals with the same $n_1$ and $n_2$.
The inversion process to determine such coefficients can turn costly for high powers of numerators.
Moreover, all solutions for the coefficients $c$ may have in principle to be derived, though only the limited subset $c(n_1,n_2)_{0,0,k}$ is relevant for the computation. The index $k$ is eventually attached to a term with the corresponding power of $z_{12}$ in the result.

In practice, I took an alternative route.
I fixed the relevant inversion coefficients $d(n_1,n_2)_{0,0,k}^{l,m,n}$ heuristically, by comparing a sufficient number of independent reduced tensor integrals with a direct IBP reduction including $z_1$ and $z_2$ scalar products.
Since the inversion coefficients must not depend on the particular integral, just on $n_1$ and $n_2$, I evaluated a test reduction on the simplest three-loop integral topology, the triple bubble, with a bunch of different numerator combinations.
Then I compared the result with a FIRE6 reduction including $z_1$ and $z_2$ contractions. This is feasible and fast also for high momentum powers, thanks to the simplicity of the integral topology.
The two sets of results must coincide, independently for each $z_{12}$ power.

This in turn produces a linear system of equations (potentially with a high number of redundant constraints), which can be solved to compute the inversion coefficients.
Moreover, for each $z_{12}$ power, these are expected to be rational functions of the dimension $d$. With some experimentation, a rough upper bound estimate of the maximal power of $d$ in the numerator and denominator of such rational functions can be established.
Let's say that these powers combined have upper bound $n$ for given $n_1$ and $n_2$.
Then, one just needs to evaluate the numeric coefficients of such powers of $d$.
For this, the relevant systems of equations can be evaluated at $n$ independent numerical values of $d$ (or a bit more for a check), where the system does not loose rank. This could be integers, or rationals, so that the final systems possess only numeric (rational) coefficients and can be inverted more rapidly.
If this is the case, the IBP reduction part can also be performed directly with rational values of the dimension, instead of generic $d$, if that produces a faster process.
One trial is sufficient for restricting a sufficient set of integrals to be IBP reduced, that supplies enough independent equations to invert the system. This eliminates the redundancy in the system, mentioned above for the subsequent reductions at fixed values of $d$.

This is the method that I applied and that I found the quickest for computations involving operators with spin 6 and higher.
For larger values of spins, it becomes more efficient to evaluate the inversion coefficients at the relevant integer dimension $d=4$ and then solve for $\epsilon$ corrections perturbatively (fixing the exact in $d$ expressions takes too much time and is practically useless). The needed order at two loops for such en expansion is $\epsilon^2$, given the maximal order of the possible poles of the relevant integrals emerging from Feynman diagrams.

In a few situations, one can determine analytically the inversion coefficients, for generic values of some of their parameters.
For instance, this is the case when only a single null momenta is present, i.e. $n_2=0$. 
The inversion coefficients only depend on the power of numerator momenta $n_1$.
A heuristic formula for the inversion coefficients can be derived for generic $n_1$, in terms of combinatorial factors.

As said, some parts of this process are a bit heuristic. For instance, the choice of numerator momenta for deriving inversion coefficients, was highly redundant. I have not studied what a minimal selection of numerators giving rise to independent constraints would be. The same goes for the exact upper bound of $d$ powers in the coefficients.
In practice, this entails some efficiency loss, and there is room for possible optimizations. However, for the computation I am presenting, the method worked sufficiently fast and there were other, more pressing bottlenecks.

Whence all the relevant inversion coefficients are known for a numerator with powers $n_1$ and $n_2$, the explicit tensor reduction of a generic integral can be carried out. This step generates a larger and larger number of scalar integrals, for higher and higher powers of numerator momenta.
A first optimization consists in leveraging the further symmetries of the reduction owing to the particular numerators. In general, numerators contain repeated powers of the loop momenta $k_1$, $k_2$ and $k_3$ (for the three-loop case) which introduce subsets of symmetric indices. This symmetry can be used, which reduces the number of independent contractions in \eqref{eq:pv}.
The output (containing scalar integrals yet to be reduced by IBP identities) can still be very bulky.
At that stage the remaining scalar integrals are reduced, expanded in $\epsilon$ up to the required order (2 at two loops) and substituted in the expressions for tensor integrals.

An initial implementation of this algorithm in Mathematica turned out to be too slow starting at spin 10.
Hence I performed most of the simplifications in FORM \cite{Vermaseren:2000nd}, which efficiently deals with large expressions.
For reducing scalar integrals via IBP reductions I mostly used FIRE6. However, some reductions failed around the level of complexity of a three-point function with two spin-10 operators, in the non-planar topology. I used Mincer \cite{Gorishnii:1989gt}, for dealing with such cases.
The process described above is rather roundabout, but it works far more efficiently than the approach in \cite{Bianchi:2019jpy} and allows to push the computation to higher values of the spins.
At spin 12 a few hundred thousands integrals were needed to be reduced.
That required deployment on a small cluster of around 300 cores for completing the process in a reasonable amount of time.

\section{Results and conclusions}

In the following tables the results are resumed for the two-loop corrections to three-point functions of twist-two operators with two spinning ones of spins $j_1$ and $j_2$ and polarization $l$. Such structure constants are normalized by the two-point functions of the corresponding operators. Said another way, the relevant operators form an orthonormal basis with respect to two-point functions. Namely, the coefficients multiplying the space-time conformal structure \eqref{eq:2point} are 1 for two-point functions of the same operator, and 0 otherwise.
Moreover, the ratio of each quantum correction $C_{j_1,j_2,l}^{(2)}$ with the corresponding tree-level value $C_{j_1,j_2,l}^{(0)}$ is reported. 
The two-loop corrections include a contribution proportional to $\zeta(3)$, which reads
$24|S_1(j_1)-S_1(j_2)|\, \zeta(3)$.
For brevity, I removed such a part from the results below, which are finally understood to represent the quantities
\begin{equation}
\bar{C}_{j_1,j_2,l} \equiv \frac{C_{j_1,j_2,l}^{(2)}}{C_{j_1,j_2,l}^{(0)}} - 24|S_1(j_1)-S_1(j_2)|\, \zeta(3)
\end{equation}
\begin{landscape}
\begin{table}
\centering
\begin{tabular}{cccccccc}
$l$& $\bar{C}_{0,0,l}$& $\bar{C}_{0,2,l}$& $\bar{C}_{0,4,l}$& $\bar{C}_{0,6,l}$ & $\bar{C}_{0,8,l}$ & $\bar{C}_{0,10,l}$ & $\bar{C}_{0,12,l}$\\[1mm]
 0 & 0 & 66 & $\frac{3532955}{31752}$ & $\frac{189088963}{1306800}$ & $\frac{29113728110377}{169682857344}$ & $\frac{2158130635015759060789}{11109996916822440000}$ & $\frac{435771907729880824453812721}{2036637039302139526560000}$
\end{tabular}
\end{table}
\begin{table}
\centering
\begin{tabular}{ccccccc}
$l$& $\bar{C}_{2,2,l}$& $\bar{C}_{2,4,l}$& $\bar{C}_{2,6,l}$ & $\bar{C}_{2,8,l}$ & $\bar{C}_{2,10,l}$ & $\bar{C}_{2,12,l}$\\[1mm]
 0 & 147 & $\frac{2712265}{15876}$ & $\frac{246135733}{1306800}$ & $\frac{857381969298607}{4242071433600}$ & $\frac{593007433738882813411}{2777499229205610000}$ & $\frac{454698118039581082150556101}{2036637039302139526560000}$ \\[1mm]
 1 & $\frac{111}{2}$ & $\frac{2718197}{31752}$ & $\frac{673255007}{6534000}$ & $\frac{4874130059289013}{42420714336000}$ & $\frac{1375252520779004546629}{11109996916822440000}$ & $\frac{266509253777609629958381701}{2036637039302139526560000}$ \\[1mm]
 2 & -87 & $-\frac{474107}{15876}$ & $-\frac{219497639}{32670000}$ & $\frac{158091242862743}{21210357168000}$ & $\frac{97527042413116934957}{5554998458411220000}$ & $\frac{51768653210494580614020901}{2036637039302139526560000}$
\end{tabular}
\end{table}
\begin{table}
\centering
\begin{tabular}{cccccc}
$l$& $\bar{C}_{4,4,l}$& $\bar{C}_{4,6,l}$ & $\bar{C}_{4,8,l}$ & $\bar{C}_{4,10,l}$ & $\bar{C}_{4,12,l}$\\[1mm]
 0 & $\frac{14378795}{63504}$ & $\frac{2916214006}{12006225}$ & $\frac{325587235460813}{1272621430080}$ & $\frac{5922079528930347198953}{22219993833644880000}$ & $\frac{561365563443718032464257591}{2036637039302139526560000}$ \\[1mm]
 1 & $\frac{20988115}{127008}$ & $\frac{368404919191}{1920996000}$ & $\frac{26680050546883301}{127262143008000}$ & $\frac{4952650041620262784703}{22219993833644880000}$ & $\frac{118798880329263503401350229}{509159259825534881640000}$ \\[1mm]
 2 & $\frac{9858115}{127008}$ & $\frac{55429699999}{480249000}$ & $\frac{8793803296085377}{63631071504000}$ & $\frac{3415365961547396145553}{22219993833644880000}$ & $\frac{336218394887834544824734241}{2036637039302139526560000}$ \\[1mm]
 3 & $-\frac{9823085}{127008}$ & $-\frac{38694354697}{3841992000}$ & $\frac{3075995211223931}{127262143008000}$ & $\frac{167602787595330815863}{3703332305607480000}$ & $\frac{60686065453702053232240433}{1018318519651069763280000}$ \\[1mm]
 4 & $-\frac{33859255}{63504}$ & $-\frac{15231046612}{60031125}$ & $-\frac{5281918551104911}{31815535752000}$ & $-\frac{886064155194573792349}{7406664611214960000}$ & $-\frac{183755638365180822400932709}{2036637039302139526560000}$ 
\end{tabular}
\end{table}
\begin{table}\centering
\begin{tabular}{ccccc}
$l$& $\bar{C}_{6,6,l}$ & $\bar{C}_{6,8,l}$ & $\bar{C}_{6,10,l}$ & $\bar{C}_{6,12,l}$\\[1mm]
 0 & $\frac{1830754919}{6534000}$ & $\frac{7767142939238407}{26512946460000}$ & $\frac{163201313679479030946391}{537723850774206096000}$ & $\frac{127291297917344130207411287}{407327407860427905312000}$ \\[1mm]
 1 & $\frac{4801454329}{19602000}$ & $\frac{56532609025579481}{212103571680000}$ & $\frac{759639819436841718481109}{2688619253871030480000}$ & $\frac{40073620608915066190184749}{135775802620142635104000}$ \\[1mm]
 2 & $\frac{91122352181}{490050000}$ & $\frac{11480219641330889}{53025892920000}$ & $\frac{9588187464549482960255129}{40329288808065457200000}$ & $\frac{516552348613652745354815891}{2036637039302139526560000}$ \\[1mm]
 3 & $\frac{5936713867}{65340000}$ & $\frac{132612893302991}{981960980000}$ & $\frac{2200778638960949186122999}{13443096269355152400000}$ & $\frac{937088520682121221186763101}{5091592598255348816400000}$ \\[1mm]
 4 & $-\frac{33451386031}{490050000}$ & $\frac{10709771810119}{1767529764000}$ & $\frac{662923727291126768354903}{13443096269355152400000}$ & $\frac{794400384020675231059986731}{10183185196510697632800000}$ \\[1mm]
 5 & $-\frac{40877954761}{98010000}$ & $-\frac{1327269952072459}{5891765880000}$ & $-\frac{1119484730526088189240969}{8065857761613091440000}$ & $-\frac{179097466961413259473566071}{2036637039302139526560000}$ \\[1mm]
 6 & $-\frac{78834756497}{32670000}$ & $-\frac{22289653409130383}{26512946460000}$ & $-\frac{273824872195577586641357}{537723850774206096000}$ & $-\frac{740555199209788799579842721}{2036637039302139526560000}$
\end{tabular}
\end{table}
\end{landscape}
\begin{table}\centering
\begin{tabular}{cccc}
$l$& $\bar{C}_{8,8,l}$ & $\bar{C}_{8,10,l}$ & $\bar{C}_{8,12,l}$\\[1mm]
 0 & $\frac{22704102808747603}{70701190560000}$ & $\frac{1783416528737094358181653}{5377238507742060960000}$ & $\frac{7328208103787158045948346317}{21511978727628848749290000}$ \\[1mm]
 1 & $\frac{86566002046549997}{282804762240000}$ & $\frac{696844256510622981986857}{2150895403096824384000}$ & $\frac{465432258097017446899761936313}{1376766638568246319954560000}$ \\[1mm]
 2 & $\frac{265016748937304477}{989816667840000}$ & $\frac{224952776474139284568427}{768176929677437280000}$ & $\frac{214598301789441142089475922519}{688383319284123159977280000}$ \\[1mm]
 3 & $\frac{1200614119876978627}{5938900007040000}$ & $\frac{5085355659157099863815449}{21508954030968243840000}$ & $\frac{359608758568933328334171885613}{1376766638568246319954560000}$ \\[1mm]
 4 & $\frac{42838323556844129}{424207143360000}$ & $\frac{56335962578633828807766253}{376406695541944267200000}$ & $\frac{31514397982642390372514409227}{172095829821030789994320000}$ \\[1mm]
 5 & $-\frac{357059515124454179}{5938900007040000}$ & $\frac{2693116666499226374227159}{150562678216777706880000}$ & $\frac{185234376440809030468641673751}{2753533277136492639909120000}$ \\[1mm]
 6 & $-\frac{362711133602400899}{989816667840000}$ & $-\frac{3033712676511763517679131}{15056267821677770688000}$ & $-\frac{156150020012746868378186296217}{1376766638568246319954560000}$ \\[1mm]
 7 & $-\frac{42825105265219741}{31422751360000}$ & $-\frac{14770837508550764976004499}{21508954030968243840000}$ & $-\frac{1236799833779991423347461560529}{2753533277136492639909120000}$ \\[1mm]
 8 & $-\frac{994642692067652671}{70701190560000}$ & $-\frac{15194325613374482381168069}{5377238507742060960000}$  & $-\frac{28822588898998607305184907023}{21511978727628848749290000}$
\end{tabular}
\end{table}
\begin{table}\centering
\begin{tabular}{ccc}
$l$& $\bar{C}_{10,10,l}$& $\bar{C}_{10,12,l}$\\[1mm]
0&$\frac{15749215446482095526927}{44439987667289760000}$ & $\frac{3291727896714092907602653231}{9057675253738462631280000}$\\[1mm]
1&$\frac{78976476506150568305851}{222199938336448800000}$ & $\frac{1678412805070059558866252441}{4528837626869231315640000}$\\[1mm]
2&$\frac{221936617587568371538721}{666599815009346400000}$ & $\frac{9620408753008166822077395353}{27173025761215387893840000}$\\[1mm]
3&$\frac{382682991742723440100673}{1333199630018692800000}$ & $\frac{114152902415962587300121940297}{362307010149538505251200000}$\\[1mm]
4&$\frac{286981761743641772075609}{1333199630018692800000}$ & $\frac{34321789080047915144640561253}{135865128806076939469200000}$\\[1mm]
5&$\frac{3236476264324378209167}{29626658444859840000}$ & $\frac{105150108827217314524556765713}{652152618269169309452160000}$ \\[1mm]
6&$-\frac{69864744286223138827091}{1333199630018692800000}$ & $\frac{1493211634156860768843828503}{54346051522430775787680000}$\\[1mm]
7&$-\frac{223404018525215783964181}{666599815009346400000}$ & $-\frac{8888956991053252742061705787}{48307601353271800700160000}$\\[1mm]
8&$-\frac{692616626682932642247029}{666599815009346400000}$ & $-\frac{242450965763881276763912033011}{407595386418230818407600000}$\\[1mm]
9&$-\frac{1127489007711542666224859}{222199938336448800000}$ & $-\frac{172363113202074410392403514227}{90576752537384626312800000}$\\[1mm]
10&$-\frac{4865338120964131771262467}{44439987667289760000}$ & $-\frac{112487449622452585777815726041}{9057675253738462631280000}$
\end{tabular}
\end{table}
\begin{table}\centering
\begin{tabular}{cc}
$l$& $\bar{C}_{12,12,l}$\\[1mm]
 0 & $\frac{779087620400198547996695657}{2036637039302139526560000}$ \\[1mm]
 1 & $\frac{807918950622770550727951307}{2036637039302139526560000}$ \\[1mm]
 2 & $\frac{8667960797896834108021796527}{22403007432323534792160000}$ \\[1mm]
 3 & $\frac{1593735587713458059151515741}{4480601486464706958432000}$ \\[1mm]
 4 & $\frac{61128677209271089728045411923}{201627066890911813129440000}$ \\[1mm]
 5 & $\frac{45594882715350651705471365527}{201627066890911813129440000}$ \\[1mm]
 6 & $\frac{474028369827181732982235661}{4073274078604279053120000}$ \\[1mm]
 7 & $-\frac{18282063762698606798756009831}{403254133781823626258880000}$ \\[1mm]
 8 & $-\frac{62843808176804166978683449193}{201627066890911813129440000}$ \\[1mm]
 9 & $-\frac{19774098758064379277379327887}{22403007432323534792160000}$ \\[1mm]
 10 & $-\frac{67981469157471204246987687547}{22403007432323534792160000}$ \\[1mm]
 11 & $-\frac{17273833425891636232934216909}{678879013100713175520000}$ \\[1mm]
 12 & $-\frac{2078775042310559062327953561427}{2036637039302139526560000}$
\end{tabular}
\end{table}
The results are available in a file attached to the submission.
It is an array in Mathematica where the entry at position $(i,j,k)$ is $\bar{C}_{2(i-1),2(j-1),k-1}$.

The first entry in the table for structure constants with one scalar operator is 0, as the corresponding operators are protected and there are no quantum corrections.
The rest of that table reproduces the structure constants with a single spinning operator, whose general formula was derived in \cite{Eden:2012rr} in terms of harmonic sums. This provides a first test of the correctness of the calculation.
The result in \cite{Eden:2012rr} was obtained via OPE from a four-point function of protected operators.
For structure constants with two spinning operators, a five-point function of protected operators would be needed.
An integrand expression was derived in \cite{Eden:2010zz}. It should in principle be possible to reproduce and possibly outperform the results presented here, from that angle. 
This would also constitute an important consistency check of the data presented here.
A first test that I have implemented on my results is the gauge invariance of the underlying Feynman diagram expansion, which is inherited from \cite{Bianchi:2019jpy}.
Moreover the fact that the three-point functions renormalize correctly and through the known anomalous dimensions of the relevant operators provides a further consistency check.
These checks however limit the tests to the gauge dependent part and divergent terms of the computation, respectively. A further check on the finite part comes from the expected behaviour of the transcendental terms, though their coefficients are tied to the divergent term, hence not independent. 
In conclusion, there are several indications that the results are correct. Still, an additional independent computation would be desirable.
The extension of the present computation to the next perturbative order, perhaps restricted to lower values of the spins, constitutes a possible interesting development, especially so since not much information on higher-point functions of protected operators is presently available at three loops.

\vfill
\newpage

\acknowledgments

This work was supported by Fondo de Investigaci\'on VIDCA 2020.
I also acknowledge computing at the Magi cluster at ICFM, Universidad Austral de Chile and thank its administrator Sim\'on Poblete for assistance and maintenance after several crashes. I thank the authors of \cite{Bercini:2020msp} for a conversation which stimulated this work.

\appendix

\section{Setting}
I work in ${\cal N}=4$ SYM with $SU(N)$ gauge group. The results presented in this article, that is up to second order in perturbation theory, do not exhibit any non-planar correction. Therefore I use ubiquitously the 't Hooft coupling $\lambda =\frac{g^2 N}{16 \pi ^2}$ as the perturbative expansion parameter ($g$ is the Yang-Mills coupling constant).
In a conformal field theory such as ${\cal N}=4$ SYM, the form of three-point functions is fixed thanks to conformal symmetry.
The focus of this note is on three-point functions with two spinning operators, of spins $j_1$ and $j_2$, and a third scalar operator.
Their three-point function is constrained to exhibit the general form
\begin{equation}\label{eq:3ptstructure2}
\left\langle \hat {\cal O}_{j_1}(x_1)\, \, {\cal O}_{BPS}(x_2) \,\,\doublehat{{\cal O}}_{j_2}(x_3) \right\rangle = \frac{ \displaystyle
\sum_{l=0}^{\text{min}(j_1,j_2)}\, {\cal C}_l\, \frac{\hat Y_{32,1}^{j_1-l}\, \doublehat{ Y}_{12,3}^{j_2-l}}{\left(x_{13}^2\right)^l} I_{13}^l}{|x_{12}|^{\Delta_{12,3}-j_1+j_2} |x_{23}|^{\Delta_{23,1}+j_1-j_2} |x_{13}|^{\Delta_{31,2}-j_1-j_2}}
\end{equation}
where the invariants read
\begin{align}
&\hat Y_{ij,k} \equiv Y_{ij,k}^{\mu}\, z_{1\,\mu}\qquad,\qquad Y_{ij,k}^\mu \equiv \frac{x_{ik}^\mu}{x_{ik}^2} - \frac{x_{jk}^\mu}{x_{jk}^2}\\&
I_{ij} \equiv I_{ij}^{\mu\nu}\, z_{1\,\mu}\, z_{2\,\nu} = z_{12} - 2\, \frac{\hat x_{ij}\, \doublehat{x}_{ij}}{x_{ij}^2} \qquad, \qquad I_{ij}^{\mu\nu} \equiv \eta^{\mu\nu} - 2\, \frac{x_{ij}^{\mu}x_{ij}^{\nu}}{x_{ij}^2}
\end{align}
and the following notation is used
\begin{equation}
x_{ij} \equiv x_i - x_j \qquad,\qquad \Delta_{ij,k} \equiv \Delta_i+\Delta_j-\Delta_k
\end{equation}
The vectors $z_1$ and $z_2$ are two null vectors ($z_i^2=0$) onto which the Lorentz tensor structure of the operators is projected, in such a way to automatically form the symmetric traceless combinations required for the correct representations of the operators.
Since there are two independent operators, two null vectors are used, which give rise to a non-trivial invariant $z_{12}\equiv z_1\cdot z_2$, parameterizing the various polarizations of the three-point function.

The coefficients ${\cal C}_l$ are $\text{min}(j_1,j_2)+1$ independent structure constants, which are functions of the coupling constant $g^2$ and the rank of the gauge group $N$, through the t' Hooft coupling $\lambda$ since, as mentioned, no color-subleading corrections appear up to two loops.
The perturbative expansion of structure constants is thus
\begin{equation}
{\cal C}_l = {\cal C}_l^{(0)} + {\cal C}_l^{(1)}\, \lambda + {\cal C}_l^{(2)}\, \, \lambda^2 + \dots
\end{equation}
As mentioned in the main text, a space-time integral is considered of \eqref{eq:3ptstructure2}, which streamlines the extraction of the structure constant.
The (dimensionally regulated in $d=4-2\epsilon$) integral of the right-hand-side of \eqref{eq:3ptstructure2} can be computed for general spins:
\begin{align}\label{eq:integrated}
& \int d^{4-2\epsilon}x_2\,\, \left\langle \hat {\cal O}_{j_1}(x_1)\, \, {\cal O}_{BPS}(x_2) \,\,\doublehat{{\cal O}}_{j_2}(x_3) \right\rangle = 
\\&~~~~~
=-\frac{\pi^2\, \hat x_{13}^{\, j_1}\, \doublehat{x}_{13}^{\,j_2}}{\epsilon\, \left(x_{13}^2\right)^{1+j_1+j_2}}\, \sum_{l=0}^{\text{min}(j_1,j_2)}\, \sum_{k=l}^{\text{min}(j_1,j_2)}\, (-2)^{k-l} \left(\begin{array}{c}k\\ l\end{array}  \right)\, {\cal C}_k\, \left(\frac{x_{13}^2\, z_{12}}{\hat x_{13}\, \doublehat{x}_{13}}\right)^l + O(\epsilon^0)\nonumber
\end{align}
The structure constants obtained via this method are not normalized in the standard form yet. The normalization is fixed in such a way that the two-point function of the relevant operators are orthonormal, which is a canonical basis of operators in a conformal field theory.
In this note I consider three-point functions of twist-two operators ${\cal O}_{j}$ of spin $j$, whose bare expression reads schematically
\begin{equation} 
O^j \equiv \Tr (D^{k}X D^{j-k} X) + \dots
\end{equation}
where $X$ is one of the complex scalars of ${\cal N}=4$ SYM and $D$ are covariant derivatives.
The correct spin representation can be attained contracting the covariant derivatives on the light-cone via null vectors $z_i$, giving rise to 
\begin{equation} 
\hat O^j \equiv \Tr (z_1\cdot D)^{k}X\, (z_1\cdot D)^{j-k} X) + \dots \qquad,\qquad
\doublehat{\bar{\mathcal{O}}} ^k(x)  \equiv \Tr ( (z_2\cdot D)^{k}\bar{X}\, (z_2\cdot D)^{j-k}\bar{X}) + \dots
\end{equation}
The scalar fields are chosen with such flavours that the three-point function is non-vanishing.
Such operators mix non-trivially under renormalization but in a closed form, that is they mix with operators of the same $sl(2)$ sector.
For twist-two operators mixing only occurs with descendants $\partial^k O^{j-k}$ of lower spin operators with the same total spin $j$. For each even value of the spin there is a single primary operator (odd spin operators are all descendants).

After re-normalization, an eigenbasis of operators for the dilatation operator is selected, whose operators are finally the conformal ones appearing in \eqref{eq:3ptstructure2}.
Their two-point function is fixed thanks to conformal symmetry
\begin{equation}\label{eq:2point}
\left\langle \hat {\cal O}^j(x_1) \doublehat{\bar{\mathcal{O}}} ^k(x_2) 
\right\rangle = C(g^2,N)\, \delta^{jk} \frac{I_{12}^j}{|x_{12}|^{2\Delta}}
\end{equation}
where $\Delta$ is the conformal dimension of the operator, including quantum corrections. The tensor structure is encapsulated in the quantity $I$.
The null vectors $z_1$ and $z_2$ are in principle distinct for the two operators, but they may be set equal to simplify the computation, without affecting the computation of the coefficient $C$. The latter is an overall normalization, depending on the perturbative definition of the operators themselves.
This is chosen canonically in such a way that the operators form an orthonormal basis. In dimensional regularization this includes subleading in $\epsilon$ corrections as well, to the required order.
The structure constants for such a basis of operators are those appearing in the results.

\bibliographystyle{JHEP}

\bibliography{biblio2}

\end{document}